\documentclass[10pt]{iopart}

\usepackage{iopams}  
\usepackage{epstopdf}
\usepackage{graphicx}

\usepackage{changes}
\usepackage{enumerate}

\begin{document}

\title[Activation of MHD reconnection on ideal timescales]{Activation of MHD reconnection on ideal timescales}

\author{S. Landi $^{1,2}$, E.~Papini $^3$, L. Del Zanna $^{1,2,4}$, A. Tenerani $^5$,  F. Pucci $^6$}

\address{$^1$ Dipartimento di Fisica e Astronomia, Universit\`a di Firenze, Largo E. Fermi 2, I-50125 Firenze, Italy}
\address{$^2$ INAF - Osservatorio Astrofisico di Arcetri, Largo E. Fermi 5, I-50125 Firenze, Italy}
\address{$^3$ Max-Planck Institut f\"ur Sonnesnsystemforschung, J. von Liebig Weg 3, D-37077 G\"ottingen, Germany}
\address{$^4$ INFN - Sezione di Firenze, Via G. Sansone 1, I-50019, Sesto F.no (Firenze), Italy}
\address{$^5$ EPSS - University of California, Los Angeles CA 90095, USA}
\address{$^6$ Dipartimento di Fisica, Universit\`a di Roma-Tor Vergata, Via della Ricerca Scientifica 1, I-00133 Roma, Italy}
\ead{slandi@arcetri.astro.it}
\vspace{10pt}
\begin{indented}
\item[]\today
\end{indented}

\begin{abstract}
Magnetic reconnection in laboratory, space and astrophysical plasmas is often invoked to explain explosive energy release and particle acceleration. However, the timescales involved in classical models within the macroscopic MHD regime are far too slow to match the observations. Here we revisit the tearing instability by performing visco-resistive two-dimensional numerical simulations of the evolution of thin current sheets, for a variety of initial configurations and of values of the Lunquist number $S$, up to $10^7$. Results confirm that when the critical aspect ratio of $S^{1/3}$ is reached in the reconnecting current sheets, the instability proceeds on ideal (Alfv\'enic) macroscopic timescales, as required to explain observations. Moreover, the same scaling is seen to apply also to the local, secondary reconnection events triggered during the nonlinear phase of the tearing instability, thus accelerating the cascading process to increasingly smaller spatial and temporal scales. The process appears to be robust, as the predicted scaling is measured both in inviscid simulations and when using a Prandtl number $P=1$ in the viscous regime.
\end{abstract}

%
%
\submitto{\PPCF}
%
%
\ioptwocol

\section{Introduction}
Most of the explosive events observed in the diverse astrophysical environments require the sudden release of the energy contained in magnetically dominated plasmas.
 Fast reconnection provides the mechanism by which the magnetic energy is channeled into heat and particle acceleration with timescales comparable to the ideal timescale $\tau_{\rm A}=L/c_{\rm A}$ (here $L$ is the characteristic length-scale of the magnetic structure and $c_{\rm A}$ the Alfv\'en speed) and much shorter than diffusion timescale $\tau_{\rm D}=L^2/\eta$ ($\eta$ being the magnetic diffusivity).
 
Sweet Parker (SP) models \cite{Sweet_1958,Parker_1957} and the linear tearing instability of a current sheet (CS) \cite{Furth_al_1963} based on the resistive magnetohyrodymanics (MHD) regime predict reconnection timescales which are too slow to explain bursty phenomena such as solar flares, magnetic substorms, and sawtooth crashes in tokamak \cite{Yamada_al_2010}. Indeed the reconnection rate in both models is very low (proportional to $S^{-1/2}$ where $S=\tau_{\rm D}/\tau_{\rm A}$ is typically $\gg 1$ in laboratory and astrophysical plasma) unless non-MHD effects are taken into account \cite{Birn_al_2001,Birn_al_2005,Cassak_Shay_2012}. 

\begin{figure*}
\begin{center} 
  \includegraphics[scale=0.50]{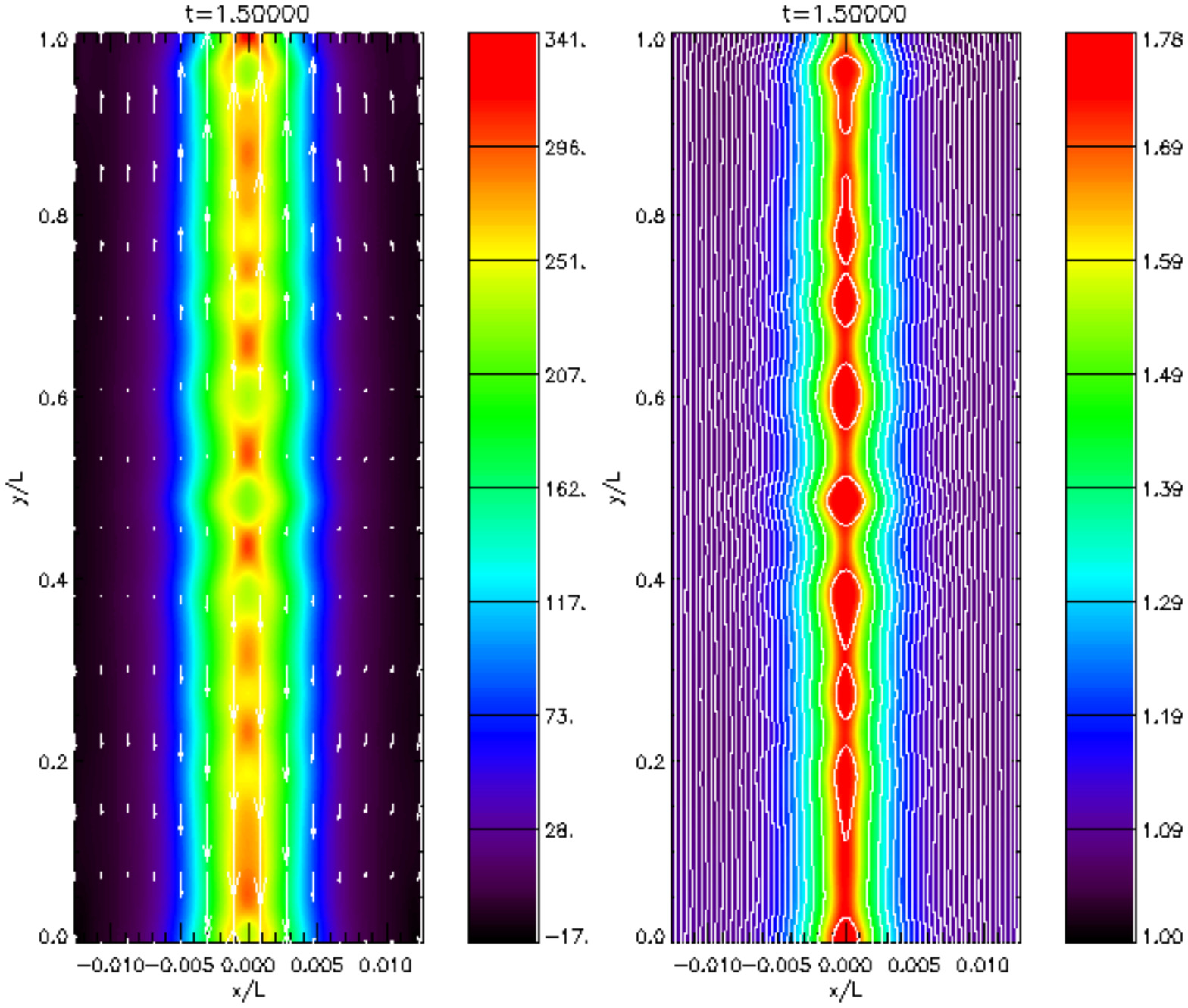} 
  \includegraphics[scale=0.35]{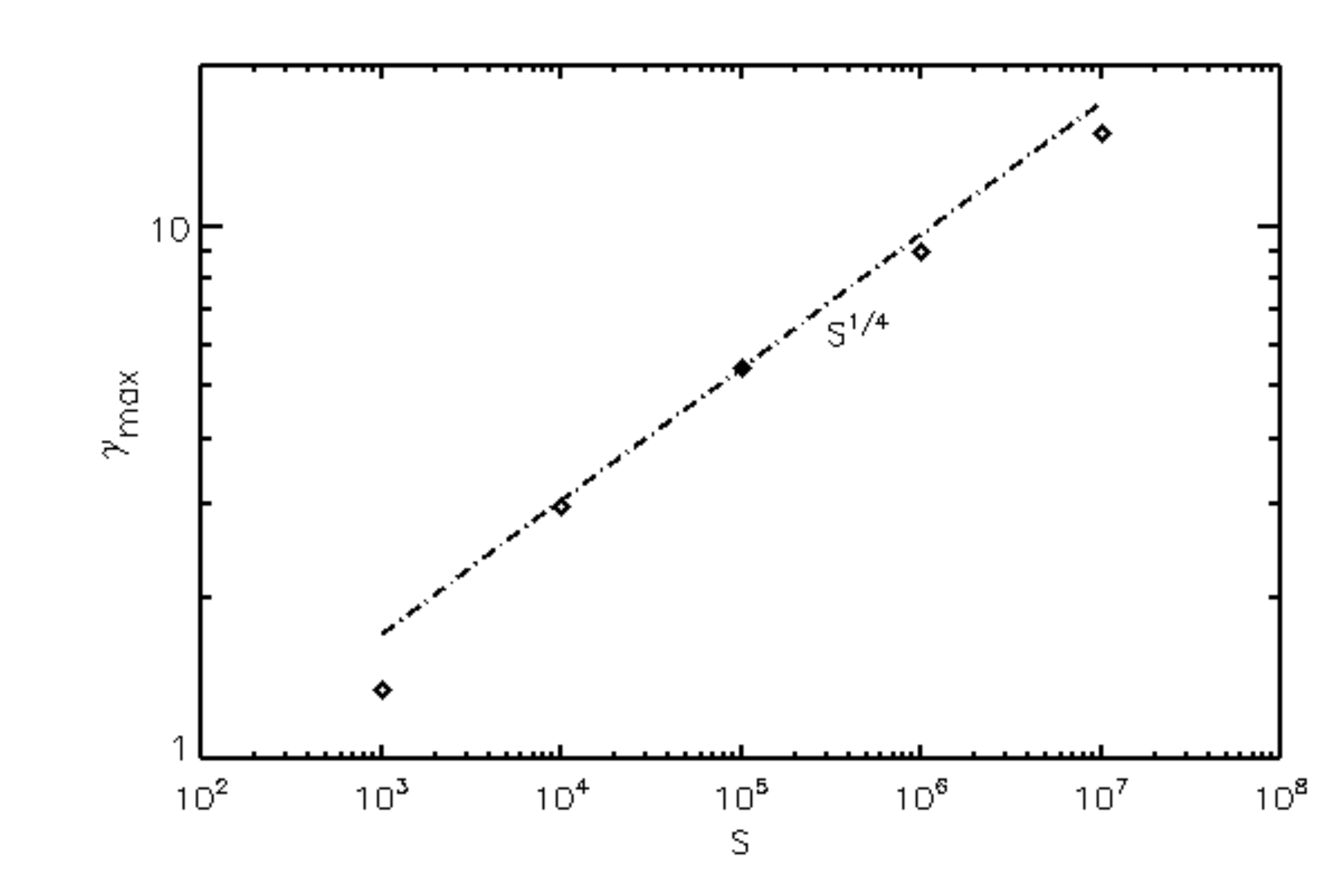}
  \includegraphics[scale=0.35]{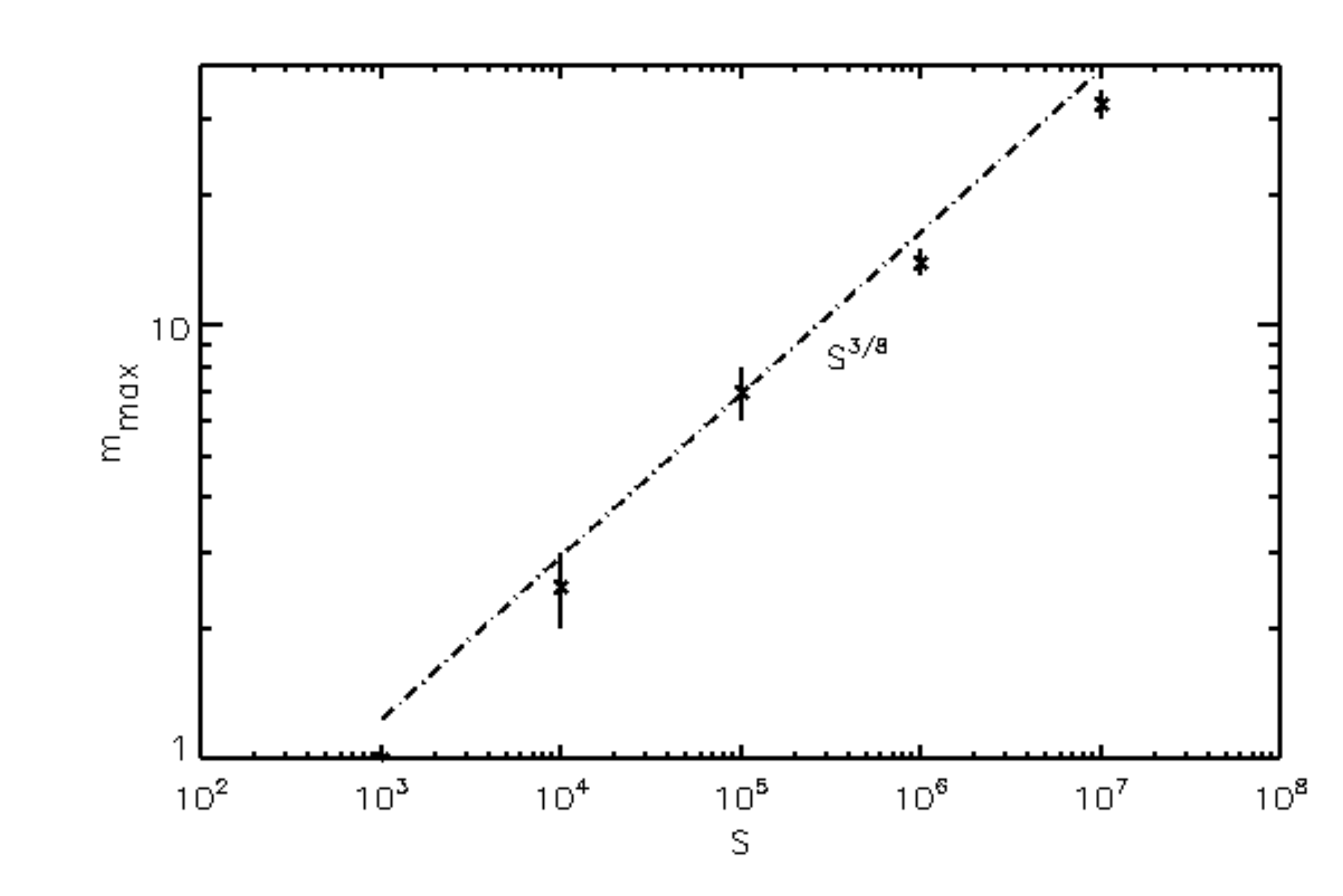} 
\end{center}
\caption{Top: An example of the plasmoid instability for a Sweet-Parker current sheet for $S=10^5$: current density intensity (colours) and velocity (arrows) on the left panel, and density (color) and magnetic field lines on the right. Bottom: Max growth rate (left) and wave-number (right) for a set of simulation of the plasmoid instability in a SP current sheet.}
\label{fig_SP}
\end{figure*}

It has been recently realized that even in a magnetofluid approach, provided that $S\gg 1$, the tearing mode can become very fast. In particular SP current sheets of aspect ratio $L/a = S^{1/2}$ ($L$ being the length and $a$ the width) were shown both theoretically and numerically to be tearing unstable with growth rate $\gamma \tau_{\rm A} \sim S^{1/4}$ \cite{Biskamp_1986,Loureiro_al_2007,Samtaney_al_2009,Loureiro_al_2009,Uzdensky_al_2010,Comisso_Grasso_2016}. The non linear evolution is charcterized by plasmoid chains with an increasing number of magnetic islands $\propto S^{3/8}$ and a reconnection rate almost independent on $S$ \cite{Lapenta_2008,Bhattacharjee_al_2009,Cassak_Shay_2009,Cassak_al_2009,Huang_Bhattacharjee_2010,Cassak_2011,Lapenta_Bettarini_2011b,Cassak_Shay_2012,Loureiro_Uzdensky_2016}.
An example of a 2D compressible MHD simulation of the SP current sheet instability  is reported in  Fig.~\ref{fig_SP} (top panels) together with the maximum growth rate and wave number as function of $S$ (bottom panels).

Resistive instabilities with growth rates scaling as a positive power of $S$ pose the problem of the existence of very thin current sheets when $S\gg 1$ since there is no ideal dynamical scale able to build up such current sheet faster than the instability itself \cite{Pucci_Velli_2014}. Magnetohydrodynamics models of magnetic closed region show that the current sheet thickness decreases reaching (potentially) the SP thickness on Alfv\'en time-scales \cite{Rappazzo_Parker_2013,Rappazzo_2015} but this process is probably stopped when the instability grows with the same time-scale.
Moreover from a conceptual point of view, an instability which grows faster as S increases, in the limit of $S \rightarrow\infty$ brings to the paradox of a infinite reconnection rate in a ideal MHD plasma, incompatible with the  "frozen-in" condition for magnetic field lines, that makes reconnection impossible (see however \cite{Eyink_Aluie_2006} for a discussion of the breakdown of the Alfv\'en theorem in ideal plasma flows).

The issue can be solved by considering a current sheet with a generic aspect ratio $a/L\propto S^{-\alpha}$ and, normalizing the tearing instability growth rate in term of the macroscopic length $L$, one has   $\gamma\propto \tau_{\rm A}^{-1} S^{-1/2+3/2\alpha}$ \cite{Bhattacharjee_al_2009,Pucci_Velli_2014}. With the SP scaling ($\alpha=1/2$) we recover the growth rate increasing as $S^{1/4}$ \cite{Loureiro_al_2007,Comisso_Grasso_2016}, while the aspect ratio $\alpha=1/3$ separates growth rates increasing to those decreasing with $S$. In this limiting case \cite{Pucci_Velli_2014} have shown that the instability is independent from $S$ with a growth rate of $\sim 0.62\tau_{\rm A}^{-1}$. 

The analytical  study of \cite{Pucci_Velli_2014}, extended to viscous incompressible flow by \cite{Tenerani_al_2015a}, have been studied numerically by \cite{Landi_al_2015,Tenerani_al_2015b,DelZanna_al_2016a,Tenerani_al_2016a}. More recently a model of such fast current sheet destabilzation has been proposed in the context of  Relativistic MHD \cite{DelZanna_al_2016b}. The scaling arguments of MHD reconnection in ideal timescales have been extended also to kinetic scales \cite{DelSarto_al_2016}.

\begin{figure}[tb]
\begin{center} 
  \includegraphics[scale=0.43]{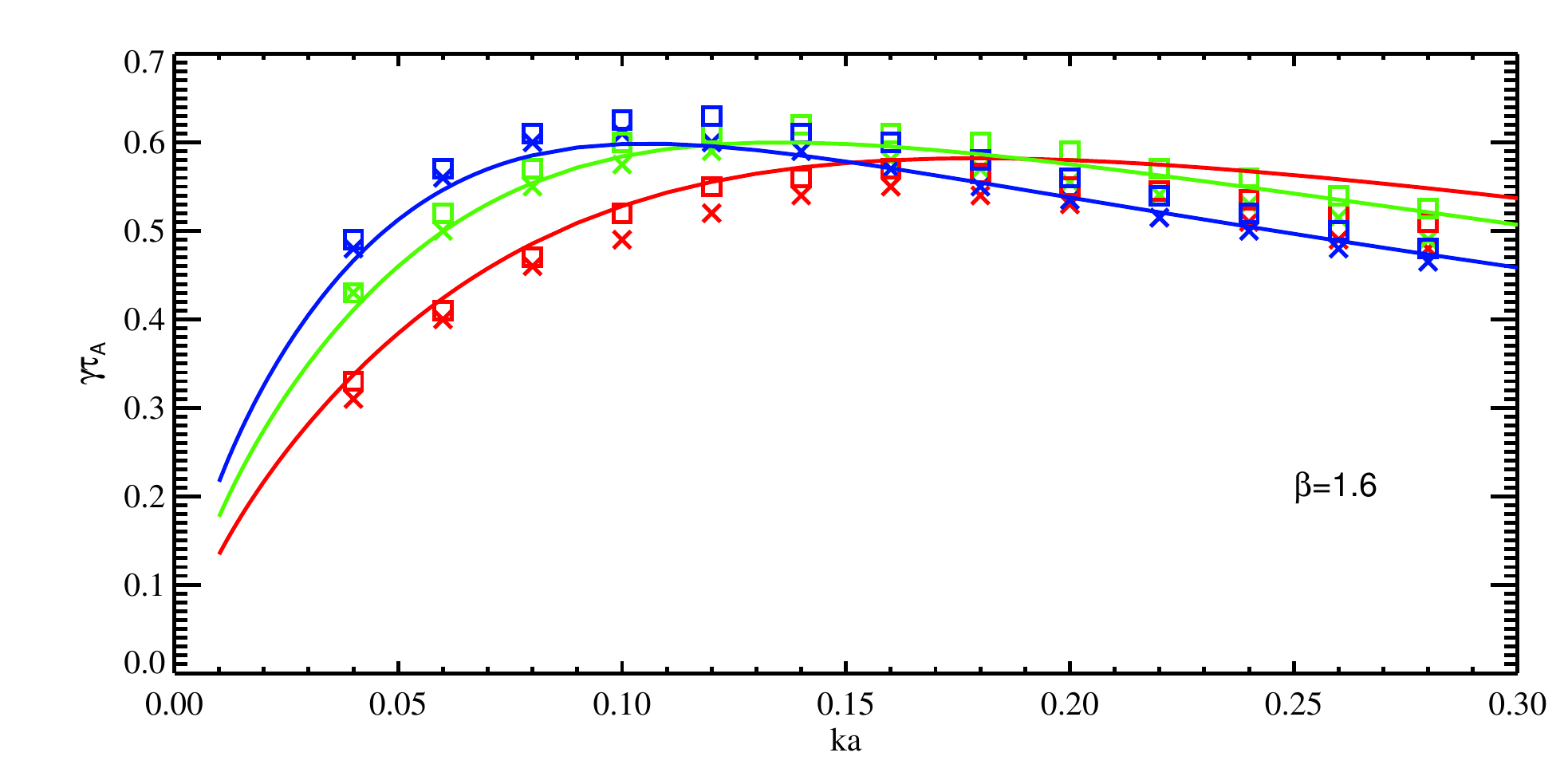} 
\end{center}
\caption{An example of the instability dispersion relation (growth rate as a function of $k$, normalized against $a^{-1}=S^{1/3}$) for different values of the Lundquist numbers (red color: $S=10^5$; green color $S=10^6$; blue color: $S=10^7$). Solid lines are the theoretical expectations, symbols are for numerical results (crosses: pressure equilibrium (PE); squares: Force-Free equilibrium (FFE), configurations). Simulations refer to a $\beta=1.6$ case (adapted from \cite{Landi_al_2015})}
\label{fig:lindisp}
\end{figure}

\cite{Landi_al_2015} solved the resistive compressible MHD equations using a numerical code which implements pseudo-spectral methods together with fourth-order compact finite-differences and characteristics for treating the (non periodic) boundary conditions \cite{Landi_al_2005}. The equations was integrated in a rectangular box and the initial equilibrium consisted of a pressure equilibrium and a force free (magnetic field rotation) equilibrium configuration. Both regime with $\beta < 1$ and $\beta > 1$ were studied and three values of $S$ were considered: $S=10^5,~10^6$, and $10^7$. The linear behaviour was analysed by measuring the dispersion relation obtained exciting one single mode for each simulation and was compared by those predicted in the linear incompressible limit. An example of such a study is reported in Fig.\ref{fig:lindisp} for the $\beta>1$ case.

Simulations and theory were in agreement both qualitatively and quantitatively since simulations show an increase in the growth rate from $\gamma \simeq 0.5~\tau_A^{-1}$ for $S=10^5$ to $\gamma>0.6~\tau_A^{-1}$ for $S=10^7$ as predicted (the asymptotic behaviour is expected for $S>10^8$ from linear theory \cite{Pucci_Velli_2014}). 
Moreover the $k$-vector corresponding to the maximum growth rate was observed to decrease as $S$ increases in terms of $a$ (but increasing in term of $L$). Also the eigenmodes were very well reproduced (see Fig.~2 in \cite{Landi_al_2015}).
The non linear evolution follows initially the typical path of the classical 2D tearing instability with the most unstable modes that start to merge, corresponding to an inverse cascade in the Fourier space. However it was recognized that secondary reconnection events arise with production of plasmoids chains. The relevant point was the fact that the onset of this secondary plasmoids occurred when the aspect ratio was of the order $S^{1/3}$ when measured on the "local" Lundquist number. As scales become smaller, the dynamical scales associated are smaller, leading to an explosive behavior.

Here we present a complementary study of such instability, focused rather on the non linear evolution, also including explicit viscosity effects. Moreover we adopt a different numerical code, ECHO \cite{DelZanna_al_2007,Landi_al_2008}, which, being based on high-order shock-capturing methods, is able to better follow the discontinuities arising from the strong non linearity driven by the tearing instabilities still preserving a sufficient high-order accuracy, which is necessary to treat properly small non ideal terms.

In section \ref{sec:Setup} the numerical setup and integration strategy is presented. Section \ref{sec:Results}  discuss some results arising from the simulations, focussing on the onset of the non linear regime. Finally a discussion is presented in section \ref{sec:Conclusion}. 

\section{Numerical setup \label{sec:Setup}}
We solve the visco-resistive compressible MHD equations expressed in conservative form:
\begin{equation}
\frac{\partial \rho}{\partial t}  + \nabla\cdot ( \rho \mathbf{v} ) = 0,
\label{eq:mhd1} 
\end{equation}
\begin{equation}
\frac{\partial \rho \mathbf{v}{2}}{\partial t} + \nabla\cdot \mathbf{M}  = 0
\label{eq:mhd2}
\end{equation}
\begin{equation}
\frac{\partial}{\partial t}\left(\rho \frac{\mathbf{v}^2}{2} + \frac{1}{\Gamma-1}p +\frac{\mathbf{B}^2}{2}\right)+\nabla\cdot\mathbf{G}=0
\label{eq:mhd3} 
\end{equation}
\begin{equation}
\frac{\partial  \mathbf{B}}{\partial t}  = \nabla \times (\mathbf{v}\times\mathbf{B} ) + \frac{1}{S} \nabla^2 \mathbf{B},
\label{eq:mhd4}
\end{equation}
with
\begin{equation}
M_{ij}=\rho v_iv_j + p\delta_{ij} -B_iB_j -\frac{1}{R}\frac{\partial v_j}{\partial x_i}
\end{equation}
and
\begin{eqnarray}
\mathbf{G}&=&\mathbf{v}\left(\frac{\rho}{2}\mathbf{v}^2+\frac{\Gamma}{\Gamma-1}p+\frac{\mathbf{B}^2}{2}\right)-\left(\mathbf{v}\cdot{\mathbf{B}}\right)\mathbf{B} \nonumber \\
                 &-&\frac{1}{S}\mathbf{B}\times\mathbf{J}-\frac{1}{R}\frac{\partial v_j}{\partial x_i} v_j~.
\end{eqnarray}
Here ${\bf \rho}$ is the mass density, ${\bf v}$ the velocity, $p$ the gas pressure, and $\mathbf{B}$ the magnetic field with $\Gamma=5/3$ the adiabatic index.  Assuming a characteristic macroscopic length scale $L$, density $\rho_0$, and magnetic field strength $B_0$, velocities are then expressed in terms of $c_{\rm A}=B_0/\sqrt{4\pi\rho_0}$ and  time in units of $\tau_{\rm A}=L/c_{\rm A}$. With this units, magnetic diffusivity is directly expressed in terms of the inverse of the Lundquist number. Moreover a kinematic viscosity $\nu$ is present  which introduces a  Reynolds number $R=S\nu/\eta=PS$ with P the Prandtl number. 
For computational reasons we use a simplified version of the kinematic viscosity where the compressible part has been neglected.

As initial conditions we assume the force-free equilibrium (FFE)
\begin{equation}
\mathbf{B}=\tanh\left(x/a\right)\mathbf{\hat{y}}+{\rm sech}\left(x/a\right)\mathbf{\hat{z}}
\label{eq:eqffe}
\end{equation}
where $a$ is the transverse (half) width of the current sheet centered on $x=0$ and elongated in the $y$ direction. Since $B^2$ is constant for this configuration, we can assume also constant density, pressure and temperature, namely $\rho=1$, $p=\rho T = \beta/2$, where $\beta$ is the plasma beta parameter.

The MHD equations (\ref{eq:mhd1})--(\ref{eq:mhd4}) are integrated in a 2D numerical box $\left[-L_x,L_x\right]\times\left[0,L_y\right]$ with resolution $N_x$ and $N_y$ respectively.   As in \cite{Landi_al_2015},  three values of the Lundquist number have been studied: $S=10^5,~10^6$, and $10^7$. The resolution is $N_x=4096$ across the current sheet (a higher resolution along x is required due to the sharp gradients for $|x| < a$) and $N_y=512$ along it; the plasma $\beta$ is $0.5$. 

The Eq.s~(\ref{eq:mhd1})--(\ref{eq:mhd4}) are solved numerically with the ECHO (Eulerian Conservative High Order) code \cite{DelZanna_al_2007,Landi_al_2008} which reconciles shock-capturing properties with high-order spectral-like differentation schemes. It is based on the Upwind Constrained Transport (UTC) methodology
\cite{Londrillo_DelZanna_2000,Londrillo_DelZanna_2004} to  treat the solenoidal condition during the magnetic field evolution and it is able to handle the ideal and dissipative set of MHD equations by using a large set of high-order methods for reconstruction, derivation and interpolation. Periodic boundary conditions are implemented along the current sheet while zeroth-order extrapolation is applied at $|x|=20a$. 
 
\section{Results \label{sec:Results}}
\cite{Landi_al_2015} performed a set of simulations to study the linear behaviour predicted by \cite{Pucci_Velli_2014}. It was considered both a pressure equlibrium (PE) and a Force-Free equilibrium (FF) configuration. By exciting one single mode for each simulation, the growth rate of the instability was measured and compared to the expected linear behaviour. It was shown in both equilibrium configurations a very good agreement between simulations  and theory with a weak dependence on the plasma $\beta$ (see for example Fig.~\ref{fig:lindisp}). A similar analysis has been performed also for this new set of simulations (limited to the FFE equilibrium configuration and using a reduced number of modes). We found a good agreement with the linear prediction when the effects of an explicit viscosity is included \cite{Tenerani_al_2015a}. Since the goal of the present simulations was the study of the fully developed non linear phase, we prefer here to present results when an initial spectrum of modes around the maximum growth rate predicted from the linear theory is initially excited (see \cite{Landi_al_2015} for further details). This initial condition has the main advantage of increasing the coupling between modes and so to reach faster the non linear phase. Moreover, the simultaneous presence of modes with different (linear) growth rate is able to produce both a direct cascade and an inverse cascade, provided that the maximum growing mode is not the smallest one in the simulation domain \cite{Landi_al_2008}, as we did. 
\begin{figure*}
\begin{center} 
  \includegraphics[width=0.8\textwidth]{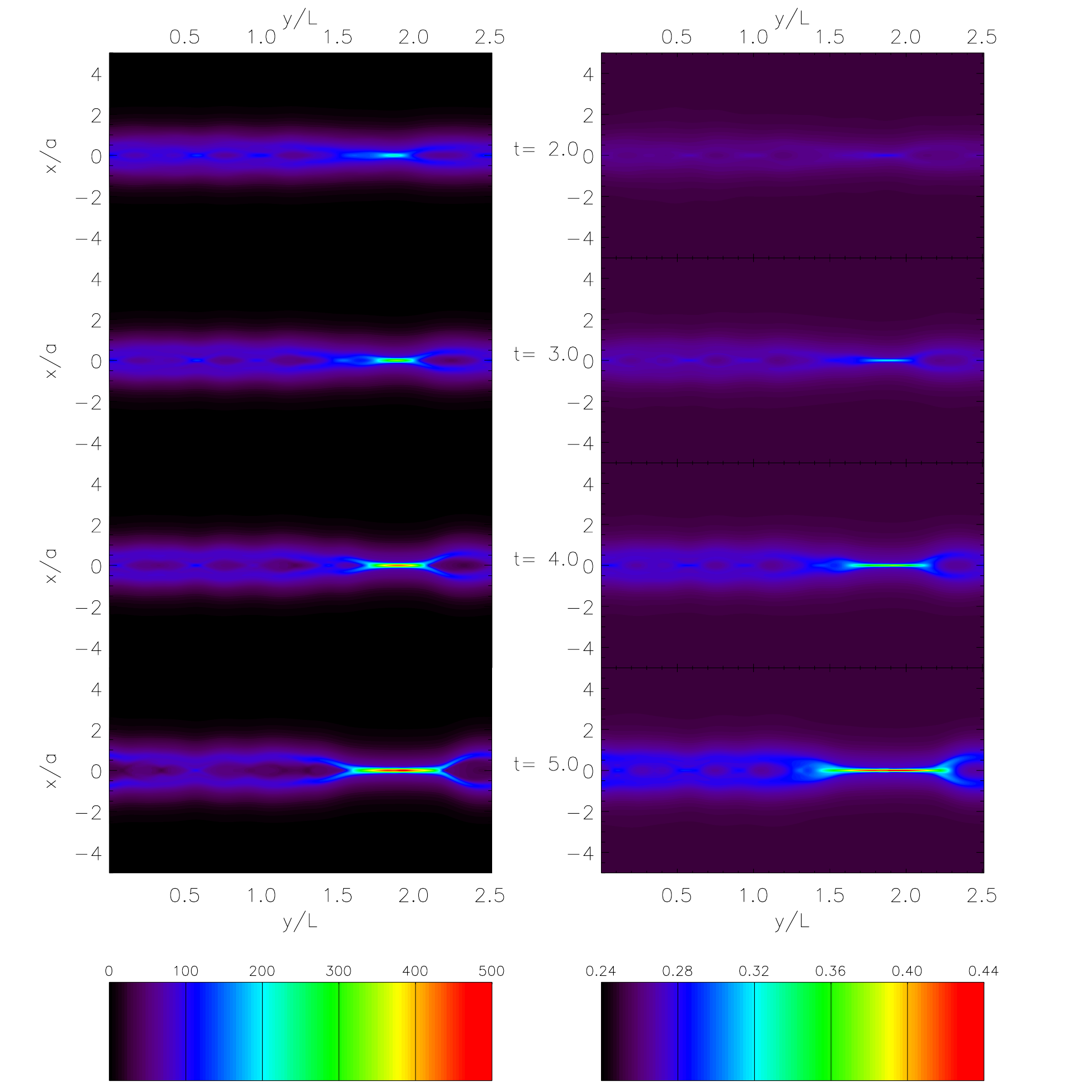} 
\end{center}
\caption{Non linear evolution of the ideal tearing instability for the case $S=10^6$ and $P=1$ . In the left panels the current density intensity is reported  as function of time from $t=2$ (top) to $t=5$. Rights panels refer to the plasma temperature at the same times.}
\label{fig_IDnl1}
\end{figure*}

\begin{figure*}
\begin{center} 
  \includegraphics[width=0.8\textwidth]{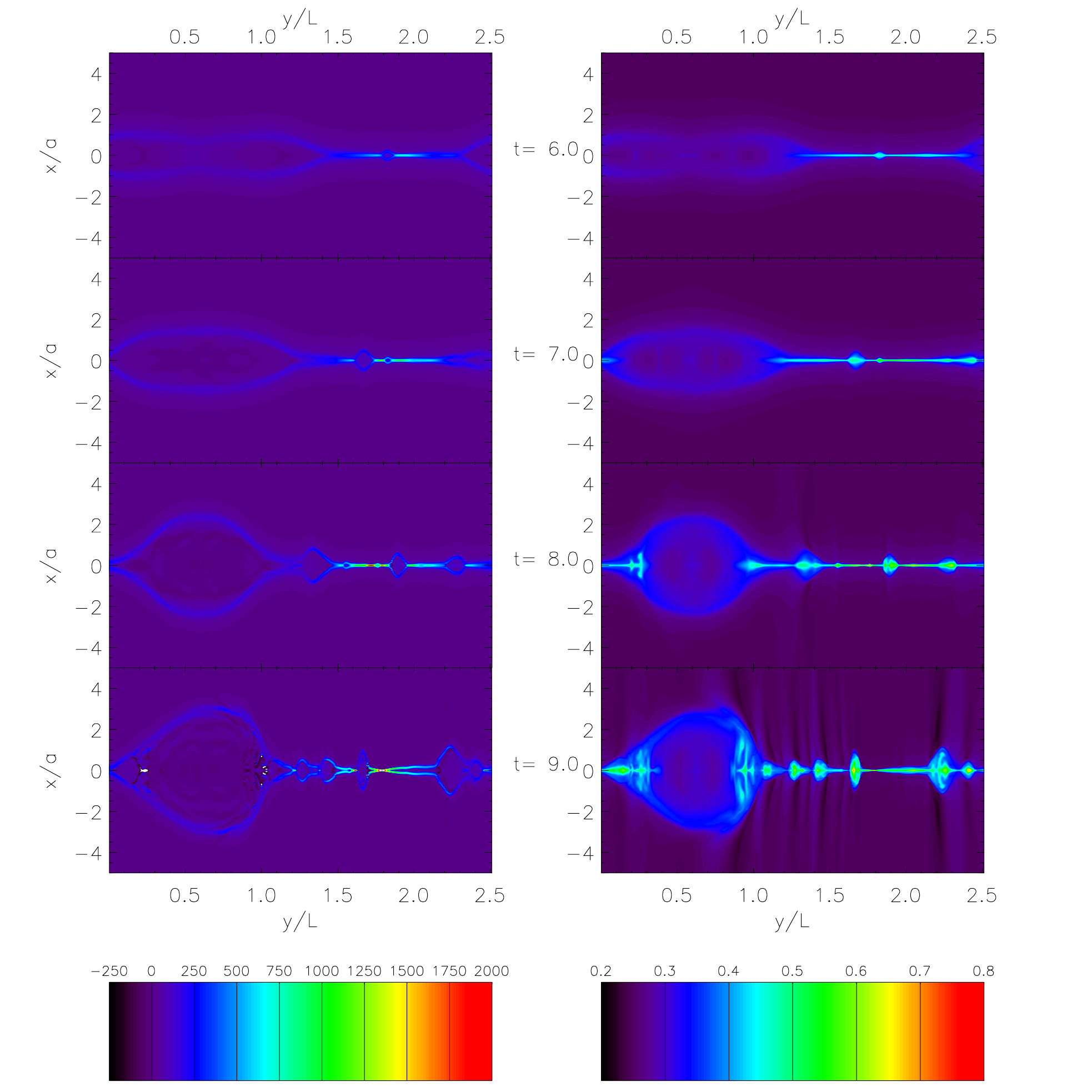} 
\end{center}
\caption{Non linear evolution of the ideal tearing instability for the case $S=10^6$ and $P=1$. In the left panels the current density intensity is reported  as function of time from $t=6$ (top) to $t=9$. Right panels refer to the plasma temperature at the same times.}
\label{fig_IDnl2}
\end{figure*}

Fig.~\ref{fig_IDnl1} and Fig.~\ref{fig_IDnl2} show 2D snapshots for selected times of a non linear simulation. Here we used $S=10^6$ and $P=1$. The simulation is extended in the interval $[-20a,20a]$ across the current sheet, although here the snapshots are zoomed in the central quarter of the simulation domain. 

Panels show the z-component of the current density $J_z$ on the left and a map of the temperature $T$ on the right. The main phases of the tearing instability are easily distinguishable. During the first phase a number of magnetic islands are forming corresponding to the wave number of the maximum growth rate in the system. The inclusion of a moderate value of an explicit viscosity ($P=1$) makes the linear phase slightly longer (the linear growth rate is reduced of about 15\%) than the pure resistive case in agreement with theoretical expectations \cite{Tenerani_al_2015a,Tenerani_al_2016a}. 

The non linear interaction among the unstable modes produces coalescence of the magnetic islands driven by the stretching of the X-neutral point associated with the most intense current sheet \cite{Malara_al_1991,Malara_al_1992b}.  At $t=7$ the fully non linear evolution is characterized by the formation of a single, large magnetic island. At the center of the most intense current sheet  newly formed plasmoids are expelled at smaller and smaller scales as typically observed during the plasmoid instability \cite{Loureiro_al_2007,Samtaney_al_2009}.  
\begin{figure*}
\begin{center} 

  \includegraphics[width=0.27\textwidth]{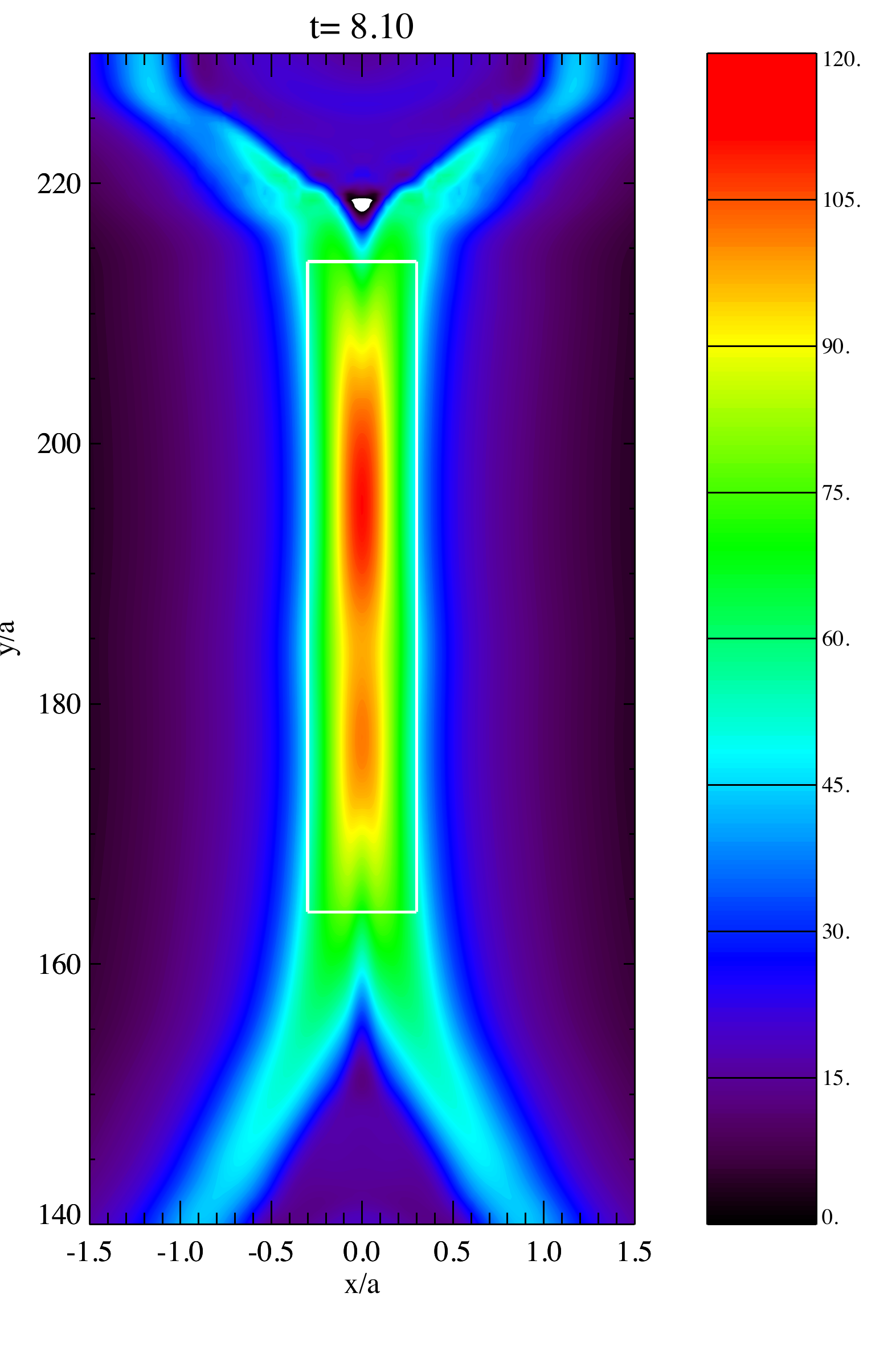} 
  \includegraphics[width=0.27\textwidth]{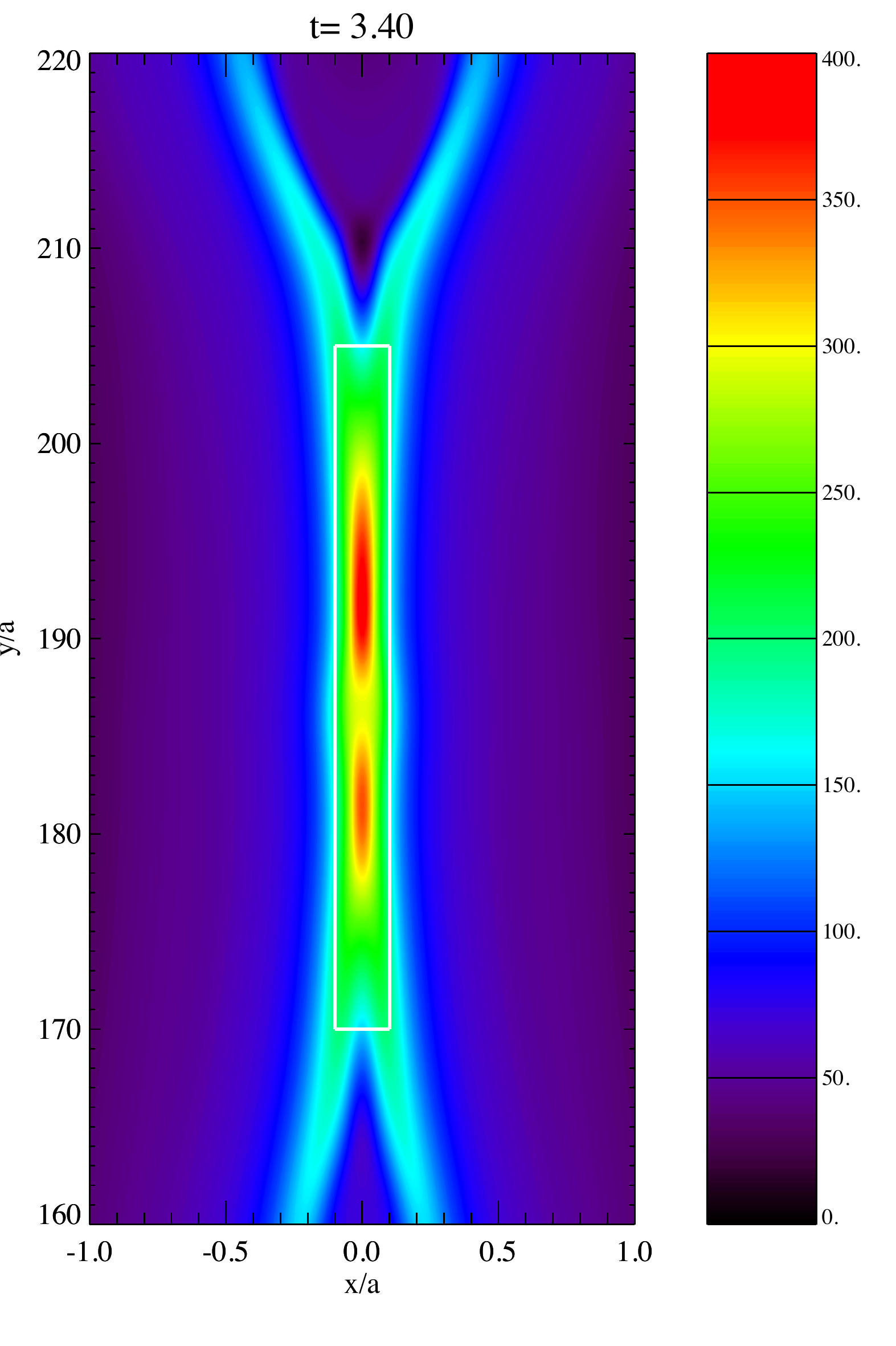} 
  \includegraphics[width=0.27\textwidth]{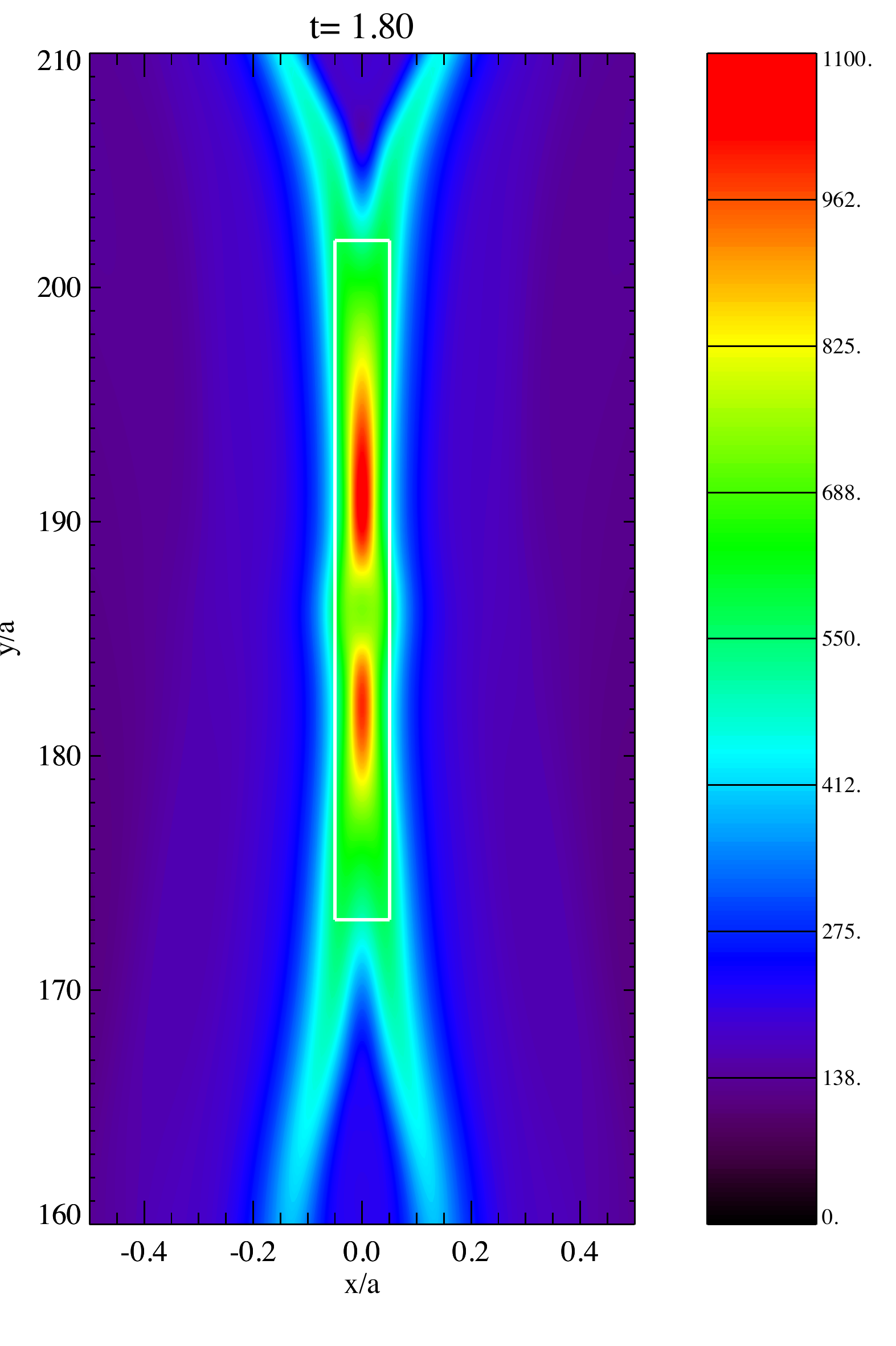} 
  \includegraphics[width=0.27\textwidth]{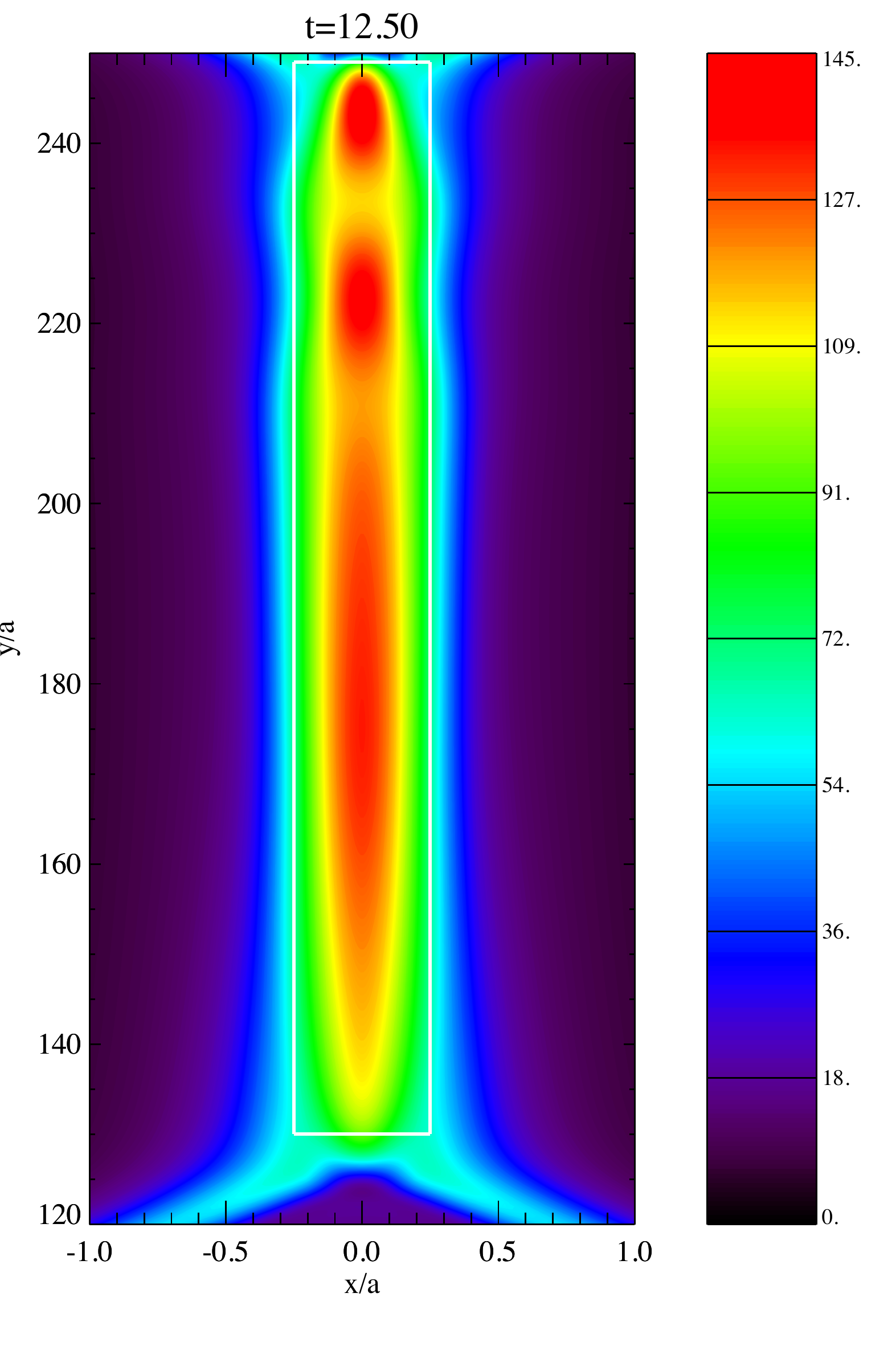} 
  \includegraphics[width=0.27\textwidth]{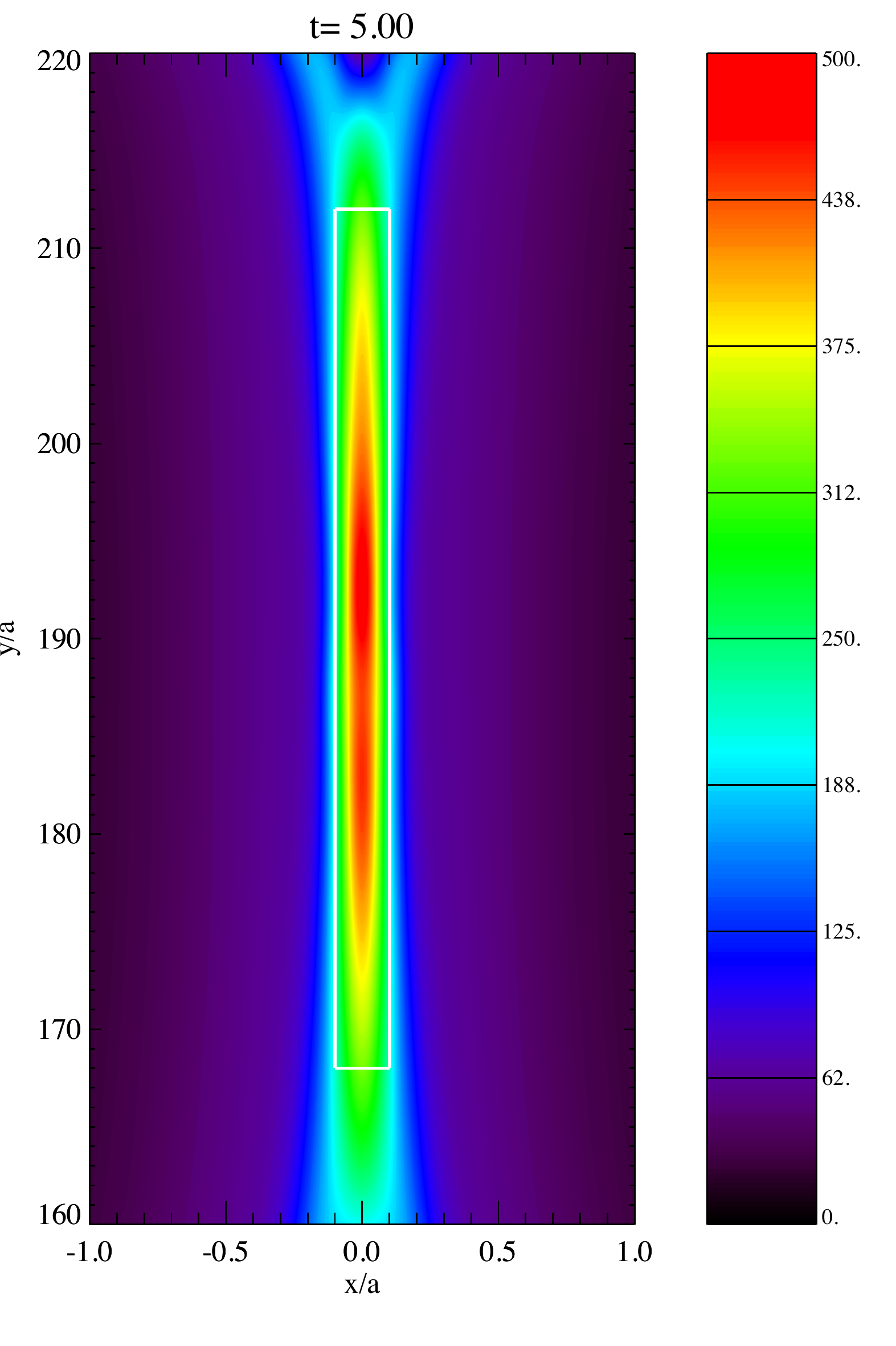} 
  \includegraphics[width=0.27\textwidth]{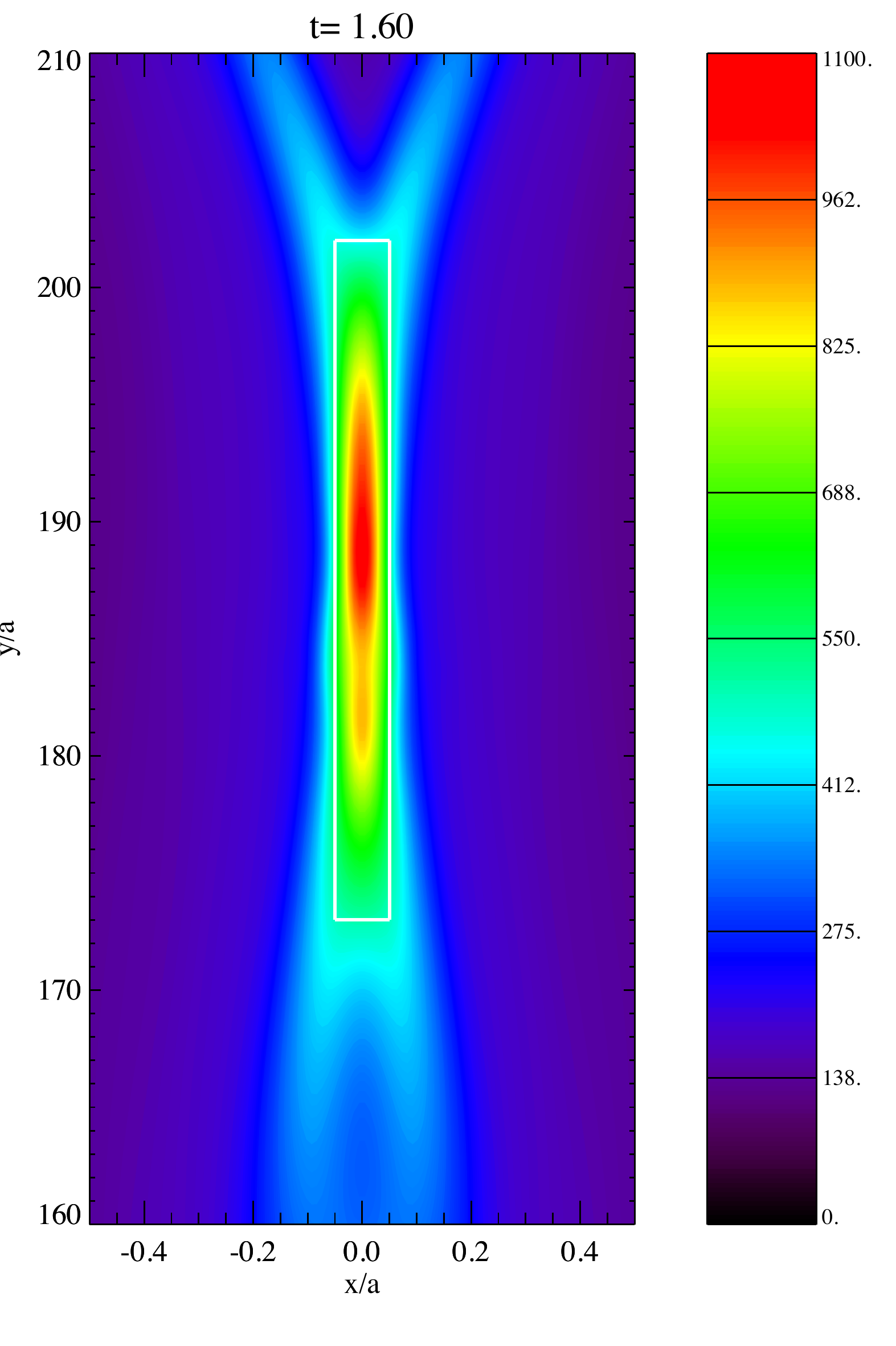} 
\end{center}
\caption{Zooms of the most prominent X-Point reconnecting regions, during the early non linear stage of the tearing instability. Colour scales correspond to the z-component of the current density intensity. From left to right runs with $S=10^5$, $10^6$, and $10^7$. The resolution adopted in these simulations is $4096$ collocation points along the cross current sheet ($x$) direction and $512$ along it ($y$ direction). An explicit viscosity is introduced with Prandtl $P=1$ for simulations in the bottom panels.}
\label{fig_IDnl3}
\end{figure*}
In \cite{Landi_al_2015} the situation right after the end of the linear phase, when the islands start to coalesce and the local current sheet has just formed and starts to further evolve, has been analysed in detail. Fig.~\ref{fig_IDnl3} reproduces the zooms around the most intense X-points formed during the linear phase for three different values of the Lundquist number ($S=10^5,~10^6$, and $10^7$) using the same format of Fig.~5 of \cite{Landi_al_2015} but for simulations with the same setup of that reported in Fig.~\ref{fig_IDnl1} and Fig.~\ref{fig_IDnl2}, with $P=0$ (top panels) and $P=1$ (bottom).  The electric current is displayed as function of $x$ and $y$ here normalized with respect to the macroscopic current sheet width $a$. We are in the phase in which, at the centre of the current sheet, the reconnection is about to take place. Defining a region where the current $J_z$ has values that are roughly half of those of the central peak (here embedded in the white box) we are able to measure the aspect ratios of these current sheets ($L^*$ and $a^*$) and measure how such values scale with the Lundquist number $S^*=(L^*/L)S$  based on the "local" macroscopic scales. Morphologically there is basically no difference between the viscid and inviscid case before the secondary current sheet undergoes the secondary explosive destabilzation. There is merely a slight elongation of the current sheet and the time where the explosion occurs is slightly delayed (according to the fact that the linear phase is longer).

\begin{figure}
\begin{center}
\includegraphics[width=0.45\textwidth]{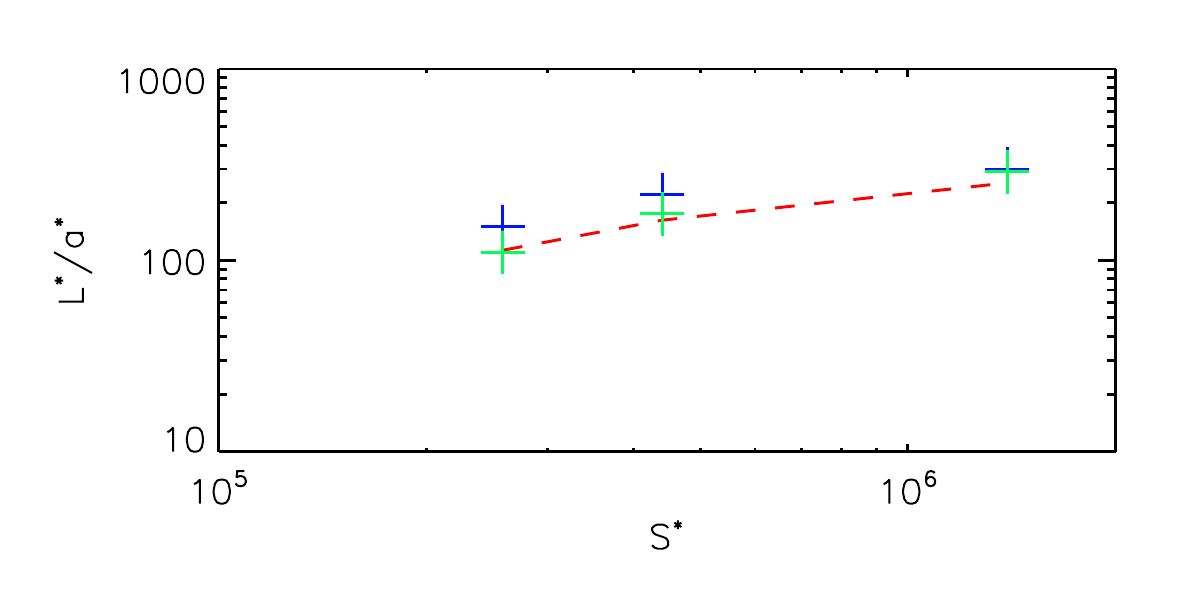} 
\includegraphics[width=0.45\textwidth]{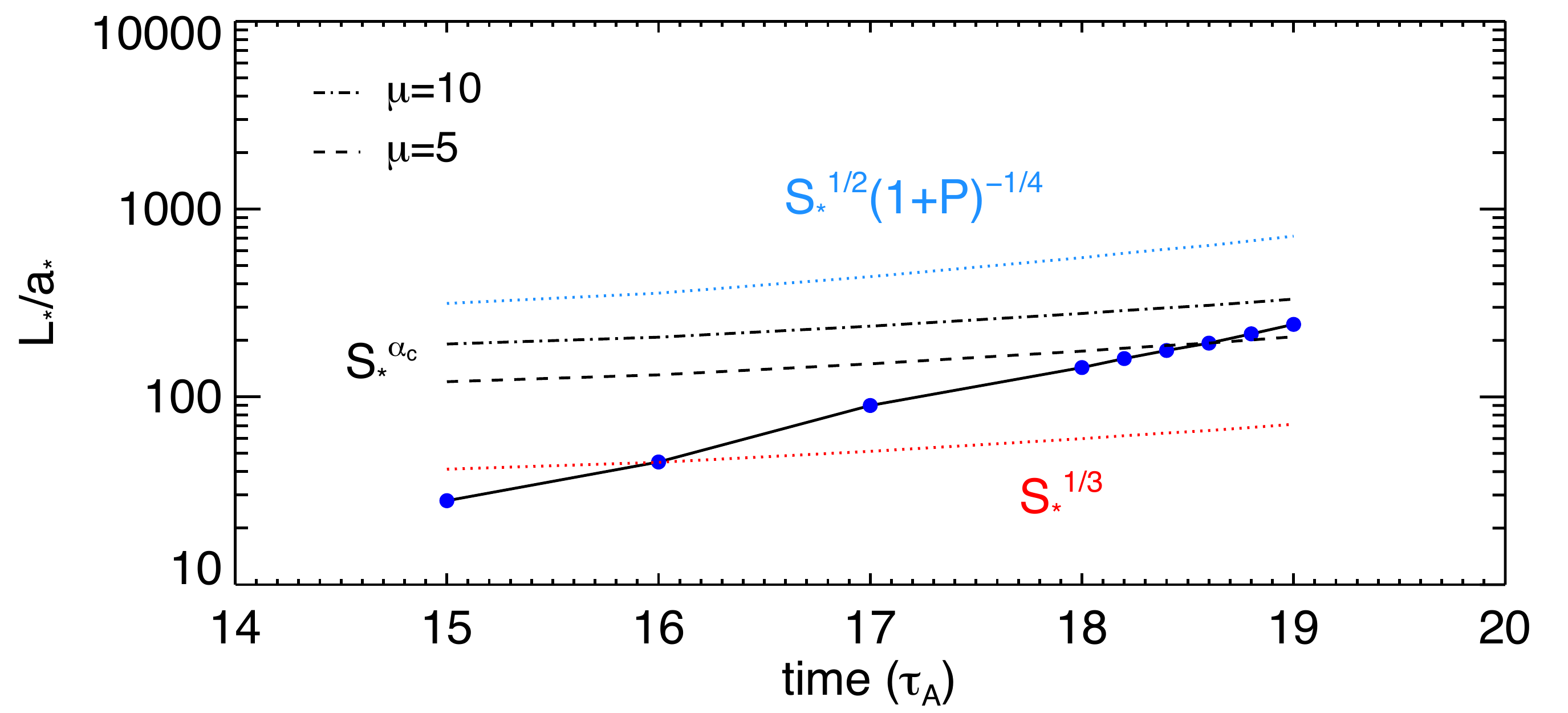} 
\end{center}
\caption{Top: Comparison of numerical results (crosses) for the local aspect ratio $L^*/a^*$ of current sheets undergoing secondary reconnection events for the current sheets shown in Fig.s~\ref{fig_IDnl1} and \ref{fig_IDnl2}. The red dashed line is the theoretical expectation $L^*/a^*\propto S^{*{1/3}}$ Green crosses refer to the case with no viscosity, $P=0$, while the blue for $P=1$.
Bottom: Blue points are the time evolution of the aspect ratio of a secondary current sheet formed after the collapse of a primary, macroscopic, current sheet. Red (blue) dotted line is the ideal (Sweet-Parker) threshold while dashed and dot-dashed lines are flow-modified aspect ratios for different plasmoid evacuation rates. Here plasmoids form at $t\sim 19$ (see \cite{Tenerani_al_2015b} for futher details). }
\label{fig_IDnl4} 
\end{figure}
In top panel of Fig.~\ref{fig_IDnl4} the local aspect ratio $L^*/a^*$ is reported as function of the local Lundquist number $S^*=(L^*/L)S$ for the simulations shown in the previous figure. Green crosses refer to simulations with no viscosity while the blue one to those with $P=1$. The scaling of the current sheet ratio with the local Lundquist number is in good agreement with a power law with index $1/3$ for both kind of simulations and in agreement with previous findings \cite{Landi_al_2015,DelZanna_al_2016a} where a different numerical approach was used. 
An analogous study has been performed by \cite{Tenerani_al_2015b} in which  the collapse of a macroscopic current sheet is studied by means of 2D visco-resistive MHD simulations. In that simulations a second current sheet forms in between two large magnetic islands and plasmoids ejections start when its aspect ratio, although larger then the one predicted by the ideal tearing, is smaller then the viscous Sweet-Parker threshold (see bottom panel of Fig.\ref{fig_IDnl4}). In that case the further thinning of the current sheet is possibly ascribed to the stabilizing effect of the outflow which reduces the growth of plasmoids until the aspect ratio is sufficiently large that the evacuation rate is insufficient to stabilize the current sheet \cite{Biskamp_1986,Biskamp_1993b,Biskamp_2000,Tenerani_al_2015b}.

\section{Conclusions\label{sec:Conclusion}}
We have presented high-resolution visco-resistive 2D MHD simulations which complement previous results of the non-linear analysis of the tearing instability of a current sheet whose inverse aspect ratio is $a/L=S^{-1/3}$, being $S$ the Lundquist number measured on the macroscopic scales and the asymptotic Alfv\'en speed.
The classical tearing instability provides a growth rate, $\propto S^{-1/2}$, which is too low to explain the explosive events often observed in nature and, 
at the same time, Sweet-Parker current sheets with $a/L\sim S^{-1/2}$ are violently unstable with growth rates diverging with $S$. A transition from slow to very fast reconnection should occur at \emph{ideal} time scales, i.~e. when the magnetic diffusion time scale becomes of the order of the ideal one. This growth rate results to be  independent from $S$ and associated to current sheets with aspect ratio of $a/L=S^{-1/3}$  \cite{Pucci_Velli_2014}.

The analytical expectations have been studied by means of linear and non linear 2D compressible MHD simulations of current sheets with aspect ratio $a/L=S^{-1/3}$ \cite{Landi_al_2015}. Single-mode simulations reproduce the dispersion relation and the tearing eigenmodes both for the classical pressure equilibrium configuration but also for the force-free configuration. The linear analytical (incompressible) analysis is weakly modified by the plasma compressibility. The non linear evolution shows the onset of the most unstable modes, the merging of magnetic islands, and secondary reconnection events. The most important result is that the evolution follows what appears to be a quasi self-similar path, with subsequent collapse, current sheet thinning, elongation, and X-point formation in the original sheet. In particular, we have verified that, even in the non linear stages of the tearing instability, the new current sheets that form locally become unstable when the inverse aspect ratio of these structures reaches the critical threshold $a/L\propto S^{-1/3}$ measured on the \emph{local}, smaller scales. As a consequence the new \emph{ideal} tearing instability starts with times occurring on shorter timescales: when time proceeds, smaller and smaller scales are reached, faster and faster reconnection arises, leading to explosive the phenomena required to explain the violent magnetic energy conversion observed in astrophysical and laboratory magnetically dominated plasmas. 
The introduction of the {\em ideal} tearing scale as the relevant one for the onset of fast collapse of the current sheet modifies models of recursive collapse \cite{Shibata_Tanuma_2001,Daughton_al_2009a} and can have strong implications in the so-called reconnection phase diagram \cite{Ji_Daughton_2011,Cassak_Drake_2013} (see discussions in \cite{Tenerani_al_2015b,Tenerani_al_2016a}). An appropriate statistical study will require very high resolution numerical simulations.

Up to now, only 2D MHD evolution of the \emph{ideal} tearing has been considered: additional dynamics is expected from kinetic effects due to the extreme thinning of the current sheet and in 3D simulations, where the onset of secondary plasmoid instability may be strongly modified by the interaction with pure 3D modes like kink and Kelvin-Helmholtz instabilities  \cite{Onofri_al_2007a,Landi_al_2008,Bettarini_al_2009,Landi_Bettarini_2012,Oishi_al_2015,Striani_al_2016}. Moreover these analyses begin with a current sheet where the critical aspect ratio has been already reached: it would be interesting to perform very high resolution simulations in which the current sheet thins, so that the aspect ratio increases with time \cite{Tenerani_al_2015b} or where is the turbulence that triggers magnetic reconnection \cite{Lazarian_al_2015}. 

\section{Acknowledgments} 
This research was conducted with high performance computing (HPC) resources provided by t by
CINECA ISCRA initiative (grant HP10CCO544).

\section*{References}

\bibliographystyle{iopart-num}


\providecommand{\newblock}{}

\end{document}